# Synthesis of Highly Anisotropic Semiconducting GaTe Nanomaterials and Emerging Properties Enabled by Epitaxy

Hui Cai, Bin Chen, Gang Wang, Emmanuel Soignard, Afsaneh Khosravi, Marco Manca, Xavier Marie, Shery Chang, Bernhard Urbaszek, and Sefaattin Tongay[*]

Hui Cai, Bin Chen, Afsaneh Khosravi, Prof. Sefaattin Tongay[*]
School for Engineering of Matter, Transport and Energy, Arizona State University, Tempe, Arizona 85287, USA
E-mail: Sefaattin.Tongay@asu.edu

Dr. Gang Wang, Marco Manca, Prof. Xavier Marie, Dr. Bernhard Urbaszek
Université de Toulouse, INSA-CNRS-UPS, LPCNO, 135 Avenue de Rangueil, 31077 Toulouse, France

Prof. Emmanuel Soignard
LeRoy Eyring Center for Solid State Science, Arizona State University, Tempe, Arizona 85287, USA

Prof. Shery Chang
John M. Cowley Center for High Resolution Electron Microscopy, Arizona State University, Tempe, Arizona 85287, USA

**Keywords**: pseudo-one dimensional materials, gallium telluride, physical vapor transport.

Pseudo-one dimensional (pseudo-1D) materials are new-class of materials where atoms are arranged in chain like structures in two-dimensions (2D). Examples to these materials include recently discovered black phosphorus (BPs)[1, 2], $ReS_2$ and $ReSe_2$ from transition metal dichalcogenides (TMDCs)[3-5], $TiS_3$ and $ZrS_3$ from transition metal trichalcogenides (TMTCs)[6, 7, 8, 9] and most recently GaTe[10]. The presence of structural anisotropy impacts their physical properties and leads to direction dependent light-matter interactions[3, 5], dichroic optical responses[8], high mobility channels[2], and anisotropic thermal conduction[11] which are especially attractive for optoelectronic and photonic applications[1, 6].

Despite the numerous reports on the vapor phase growth of isotropic TMDCs and post transition metal chalcogenides (PTMCs) such as $MoS_2$[12] and GaSe[13], the synthesis of pseudo-1D materials is particularly difficult due to the anisotropy in interfacial energy, which stabilizes dendritic growth rather than single crystalline growth with well-defined orientation[5, 14]. The growth of monoclinic GaTe has been demonstrated on flexible mica substrates with superior photodetecting performance[15]. In this work, we demonstrate that pseudo-1D monoclinic GaTe layers can be synthesized on a variety of other substrates including GaAs (111), Si (111) and c-cut sapphire by physical vapor transport techniques. High resolution transmission electron microscopy (HRTEM) measurements, together with angle resolved micro-PL (ARMP) and micro-Raman (ARMR) techniques, provide for the very first time atomic scale resolution experiments on pseudo-1D structures in monoclinic GaTe and anisotropic properties. Interestingly, GaTe nanomaterials grown on sapphire exhibit highly efficient and narrow localized emission peaks below the band gap energy, which are found to be related to select types of line and point defects as evidenced by PL and Raman mapping scans. Detailed spectroscopic studies reveal unique features of these sharp sub-band emissions, which distinguish themselves from the broad defect emissions normally found in semiconductors. It also makes the samples grown on sapphire more prominent than those grown on GaAs and Si, which demonstrate more regular properties. Overall findings offer new routes to synthesize pseudo-1D GaTe layers, provide synthesis insights and epitaxy relations, and establish their anisotropic optical performance. These unique properties open new opportunities for GaTe to find applications in novel optoelectronic devices that have been established in the PTMC family[16].

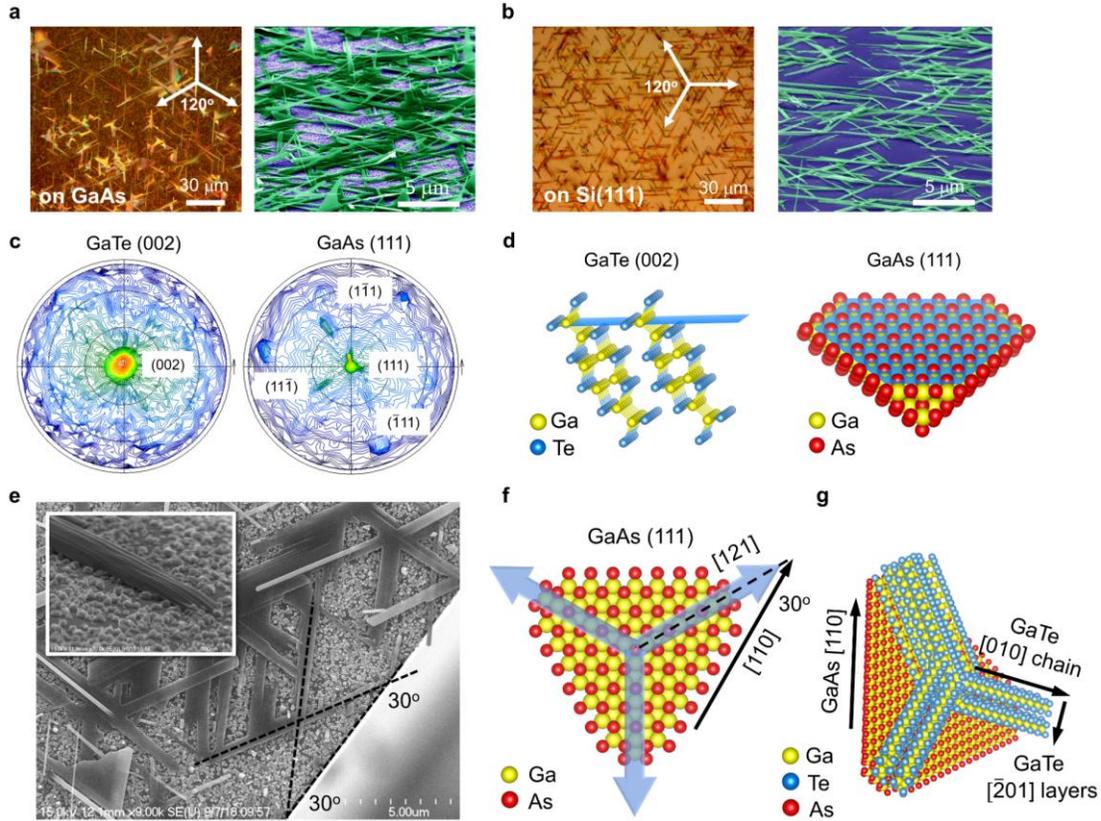

**Figure 1. a)**. Optical (left) and false colored SEM (right) images of GaTe synthesized on GaAs (111) substrate. **b)**. Optical (left) and false colored SEM (right) images of GaTe synthesized on Si (111) substrate. **c)**. Pole figure XRD spectra of the GaTe (002) plane and the GaAs {111} plane system. **d)**. The atomic structure of both planes are shown in **(c)**. **e)**. SEM image of GaTe grown at the edge of the GaAs (111) substrate. **inset**: Zoom-in view of a GaTe nanowire grown on GaAs (111). **f).** atomic structure of the GaAs (111) surface with the GaTe nanowire growth direction labeled by blue arrows. **g)**. 3D atomic model of the GaTe nanowires grown on GaAs (111).

**Synthesis and structural characteristics.** In this work, GaTe flakes were synthesized by physical vapor transport (PVT) technique in a tube furnace (**Figure S1**) using GaTe and $Ga_2Te_3$ polycrystalline powders as the source. A detailed description of the growth process can be found in the experimental section. We have performed the growth on a variety of substrates including GaAs (111), Si (111) and c-cut sapphire to understand different substrate effects and growth mechanism. Here, GaAs and Si wafers were selected in order to integrate 2D materials systems into scalable bulk semiconductors in hybrid structures, and sapphire was chosen owing to recent successful growth of 2D systems onto their highly crystalline surfaces.

As shown in **Figure 1a-b**, the morphology of the GaTe nanostructures depends strongly on the substrate type. Typical growth processes carried onto GaAs and Si yield GaTe nanowires lying flat on the surface, however some of the ribbons also grow out of plane at an acute angle with the substrate as shown in scanning electron microscopy (SEM) images in **Figure 1a-b**. These nanowires appear to be well aligned along three directions that make a fixed angle of 120º with each other, which is a sign of epitaxial growth[17]. To confirm epitaxial relationship between GaTe and GaAs, we have performed pole figure X-ray diffraction (XRD), SEM and HRTEM combined. A full XRD spectrum of the sample is shown in **Figure S2**. GaAs (111) and (222) peaks are clearly seen, confirming the (111) orientation of the GaAs substrate. As for GaTe, only two peaks are revealed-the (001) peak at 11.9º and the (002) peak at 23.8º. This suggests that the GaTe nanowires have a (002) preferred orientation on GaAs(111) as shown in **Figure 1d**. The pole figure XRD result shown in **Figure 1c** further confirms that the GaTe (002) plane is parallel to the GaAs (111) plane (**Figure 1d** and inset of **Figure 1e**).

Next, we determine the rotational relationship between the GaTe nanowires and the GaAs substrate by analyzing the angle of the nanowire growth. **Figure 1e** shows the SEM image of the GaTe nanowires grown at the edge of the GaAs wafer. It is known that the GaAs (111) wafer cracks along the [110] direction. We find that the nanowires form a 30º angle with respect to the wafer edge, which is parallel to the [121] direction of GaAs (**Figure 1f**). HRTEM image (**Figure S3**) shows that the nanowire growth direction is along the [010] chain direction of GaTe. Thus the rotational relationship between GaTe and GaAs is determined as GaTe [010]//GaAs [121]. It is worthy to note that the growth of GaTe is quite different in that instead of isotropic growth (i.e., material grows in all directions in the plane), monoclinic GaTe growth occurs preferably along the [010] chain direction. This chain-like growth is directly related to the highly anisotropic crystal structure[9] of monoclinic GaTe and will be discussed later in the article. Here, we also note that similar relations also exist for GaTe grown onto Si wafers but are not shown here for brevity. However, due to the large lattice mismatch of 13.7% between GaTe (4.15 Å of Te-Te distance along [010]) and sapphire (4.81 Å along [010]) compared to 2.0% and 7.2% on GaAs ($d_{As-As}$=4.07 Å along [110]) and Si ($d_{Si-Si}$=3.87 Å along [110]) respectively, GaTe flakes grow rather randomly on sapphire substrates as shown in **Figure 2a**. The resulting morphology varies from wires to flakes with the dimensions ranging from 10 μm to 30 μm. This random distribution of morphology and dimension may be attributed to the van der Waals epitaxy growth mechanism where the GaTe is bonded to the substrate by weak van der Waals force without forming any direct chemical bonds. Under this regime, the GaTe adatoms can transport more freely on the sapphire surface. Overall our manufacturing route enables synthesis of pseudo-1D GaTe sheets for the first time onto different substrates and is a significant progress in synthesis of pseudo-1D systems.

**Optical response of CVD deposited GaTe.** To understand how substrate type and growth characteristics influence the overall optical response from direct band gap nanomaterial GaTe, we have performed photoluminescence (PL) measurements as shown in **Figure 2b**.

We find that the GaTe flakes grown on sapphire behave quite differently from those grown on GaAs and Si as well as flakes isolated from GaTe single crystals synthesized via Bridgman method. The main PL peak located at 1.66 eV for GaTe/sapphire (red) and GaTe/GaAs (green) matches closely the fundamental direct band gap of GaTe at 1.65 eV (blue line), and is thus identified as the band edge emission ($X_0$) related to radiative recombination of photo-excited electrons and holes. The sample grown on GaAs displays an additional peak at 1.42 eV (green line) originating from the GaAs substrate ($E_g^{GaAs}$=1.4 eV). Surprisingly, however, the sample grown on sapphire (red line) shows three additional sharp (FWHM~50 meV) emission lines at 1.29 eV, 1.39 eV and 1.50 eV as pointed out by the blue arrows which are not present in exfoliated GaTe and GaAs/GaTe samples. Then the question arises: What is the origin of these additional peaks?

As evidenced by micro-Raman measurements, we safely exclude the presence of secondary phases such as $Ga_2Te_3$ or $TeO_2$ and their potential impact on the PL spectrum. As shown in **Figure 2c**, Raman spectra from sapphire/GaTe precisely match the GaTe single crystals grown by the Bridgman method, GaAs/GaTe, as well as existing literature without any additional peaks associated with other compositions and phases[18]. The Raman measurements were performed on various spots and none of them showed the possible existence of a second phase. Another direct evidence for single phase GaTe comes from our selected area electron diffraction (SAED) patterns and energy dispersive X-ray spectroscopy (EDS) mapping in large areas on a GaTe flake synthesized on sapphire (**Figure S4**). In agreement with our earlier Raman data, the SAED shows the flake is pure crystalline GaTe without any second phases. The EDS mapping also indicates a uniform distribution of Ga and Te elements. However, the SAED does suggest that the flake is poly-crystalline and is composed of two domains with different crystal orientations, which will be discussed later with the anisotropy properties. The high resolution transmission electron microscopy (HRTEM) image (**Figure 3b**) also proves the high crystallinity of the GaTe flake.

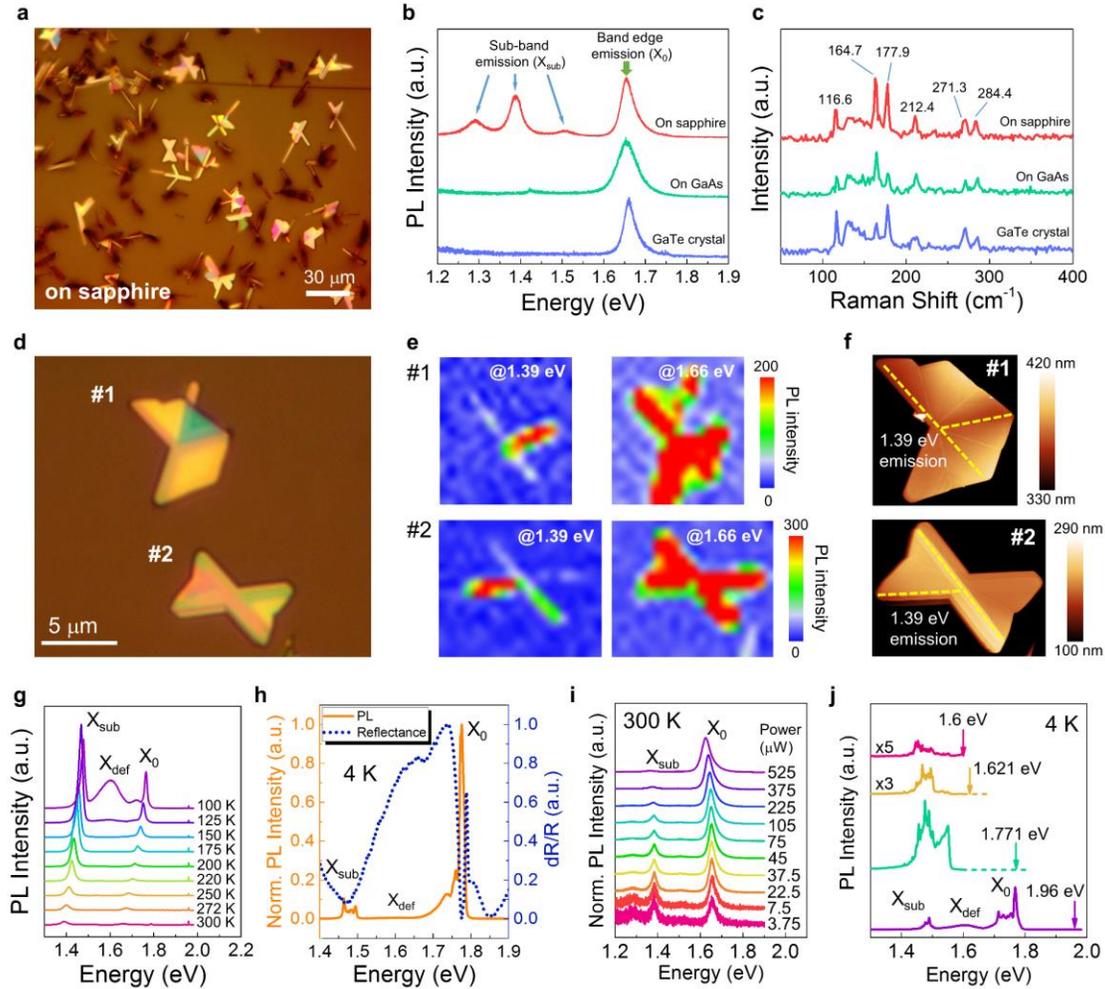

**Figure 2. a)**. Optical image of GaTe synthesized on c-cut sapphire substrate. **b)**. Photoluminescence spectra of GaTe synthesized on various substrates. **c)**. Raman spectra of GaTe synthesized on various substrates. **d)**. Optical image of two typical GaTe flakes grown on sapphire. **e)**. PL intensity mapping at 1.39 eV and 1.66 eV for the two flakes. **f)**. AFM images of the two flakes. yellow dashed lines show the position for the 1.39 eV emission. **g)**. Temperature dependent PL spectra measured at the spot with 1.39 eV emission. **h)**. PL and photo reflectance spectra measured at the spot with the intermediate band emission at 4 K. **i)**. Power dependent PL spectra

measured at the spot with the intermediate band emission at room temperature. **j).** PL spectra of the GaTe flakes excited by different laser energies at 4 K.

To elucidate on the origin of these new PL peaks, we have performed large area spatial mapping of the PL spectra on two typical flakes grown on sapphire under 488 nm laser excitation with particular focus on the two most prominent PL peaks, namely 1.39 eV ($X_{sub}$) and 1.66 eV ($X_0$). However, similar conclusions can also be drawn for other below band gap emission peaks in **Figure 2b.** Overall, PL mapping data shows that emission peak position and intensity are relatively non-uniform across the flake as shown in **Figure 2d-e**. The $X_0$ emission is observed all across the flake for both flake #1 and #2, and is particularly strong at certain boundaries and edges. The $X_{sub}$ emission line, however, is only observed at selected boundaries and edges, as shown in **Figure 2e** and **2f,** suggesting spatially localized emission. A line scan of the PL spectra across flake #2 can be found in **Figure S5**, which clearly demonstrates that the $X_{sub}$ emission is only localized at the edges.

Temperature dependent PL measurements (100-300K) in **Figure 2g** show that FWHM of $X_0$ and $X_{sub}$ both remain fairly sharp at all temperatures (15 and 22 meV, respectively). Below 150 K an additional, broad (FWHM~123 meV) peak appears at 1.55 eV. This peak is also seen at 4K as shown in Figure S6. In contrast to $X_{sub}$, this broad peak, which thermally quenches above 100K, can be attributed to randomly arranged defects with shallow confinement, such as Ga vacancies that have been studied recently[19]. High signal-to-noise photoreflectance (PR) spectra at 4K (**Figure 2h**) show that the sharp differential reflectance (dR/R) peak (blue dashed line) matches closely in energy with the $X_0$ PL emission line (orange solid) i.e. we do not observe any Stokes shift due to localization of Coulomb bound electron-hole pairs (excitons) at this energy. This clearly demonstrates that the emission comes from radiative recombination of excitons at the direct bandgap around 1.78 eV[20]. Here, we note that the blue shift of the $X_0$ emission from 1.66 eV at room temperature to 1.78 eV at low temperature is simply due to increase in optical band gap which can be described by the Varshni Law[21], the exciton binding energy of around 20 meV (bulk GaTe) is only a small correction[20]. Similarly, $X_{sub}$ peak position shifts from 1.39 eV at 300K to 1.5 eV at 4K, but the photoreflectance spectrum does not display any transition peak at 1.5 eV. The absence of reflectance transition proves that the $X_{sub}$ has a comparably low density of states and oscillator strength and is not related to any strong band-to-band transition. We argue that $X_{sub}$ originates from recombination of electron-hole pairs at localized energy levels at 1.39 eV (localized emission centers). This can be clearly seen in our power dependent PL studies: As shown in **Figure 2i** the $X_{sub}/X_0$ peak intensity ratio quickly decreases and the $X_{sub}$ peak saturates as the laser power density is increased. This behavior is another indication for a low density of states and suggests recombination of carriers spatially localized at defects to be the origin of $X_{sub}$.

Using different laser excitation energies, we provide further evidence that shows that $X_{sub}$ does not come from the band-to-band transition of a second phase but structural defects in GaTe. **Figure 2j** shows the PL spectra at excitation energies above the band gap (1.96 eV), resonant with the band gap (1.771 eV) and below the band gap (1.621 and 1.6 eV). Carriers excited resonantly at the band-gap can clearly relax very efficiently towards $X_{sub}$, which suggest that defects can capture a considerable fraction of the generated excitons. As expected from this scenario, we observe that laser excitation *below* the gap results in much lower defect related $X_{sub}$ emission as the absorption (i.e. exciton generation) is strongly diminished.

The significance of this $X_{sub}$ emission can be described as follows: (i) This emission is clearly seen at low temperature (4K) and room temperature, suggesting either strong confinement at defect sites or the formation of a mid-gap defect band. This looks similar to the intermediate band emission found in $ZnTe_{1-x}O_x$[22] and $CuIn_{1-x}Se_xS_2$[23] recently, which indicates the possibility of below bandgap absorption and intermediate band solar cells based on GaTe; (ii) The emission is spectrally very sharp with a FWHM similar to the band edge emission $X_0$, which is abnormal for a defect emission. Both observations (i) and (ii) are in clear contrast to $X_{def}$, which only appears at low temperature and is spectrally broad. Further studies have been planned on determining what kind of defects give rise to $X_{sub}$ exactly. One possibility is $O_2$ adsorption as it has been established by DFT calculations that the GaTe-$O_2$ complex can induce intermediate band states in GaTe[24].

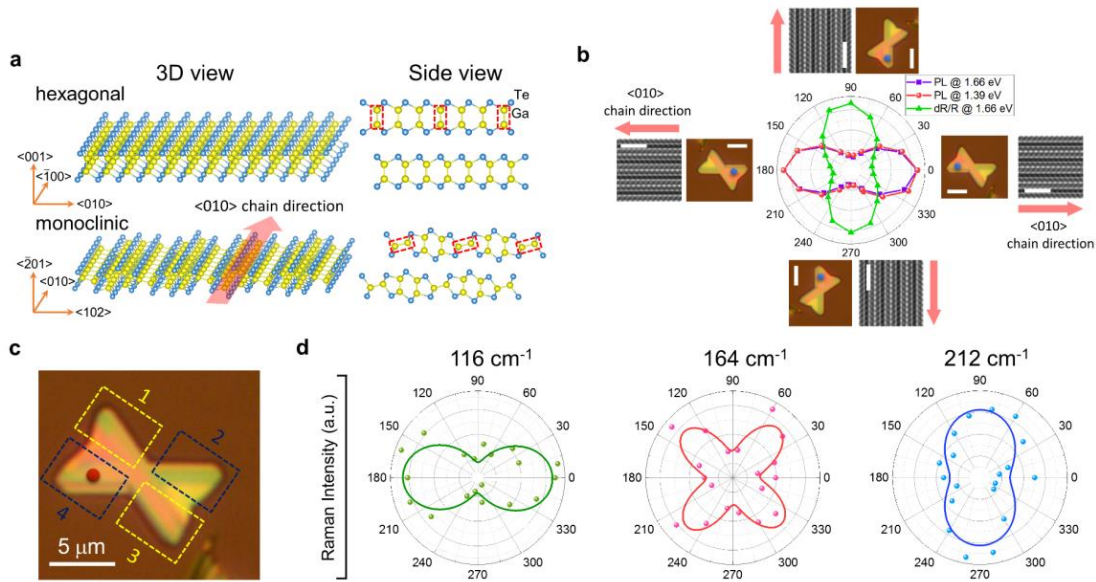

**Figure 3. a).** Schematics of the crystal structure of hexagonal and monoclinic GaTe. **b).** Angle resolved PL and reflectance intensity of the GaTe flake. The optical (scale bar = 5µm) and HRTEM images (scale bar = 2 nm) corresponding to 0º, 90º, 180º and 270º are shown around, with red arrows pointing toward the <010> chain direction. The HRTEM image is processed by inverse fast Fourier transform (i-FFT) from the original one shown in Figure S3. The blue spot shows the location where the PL and HRTEM are measured. **c).** Optical image of the same GaTe flake, with a red spot showing the location where the angle resolved Raman spectroscopy is measured. The domain structure is indicated by the rectangles. **d).** Angle resolved Raman intensity of the 116 $cm^{-1}$, 164 $cm^{-1}$ and 212 $cm^{-1}$ peaks of the GaTe flake. The solid lines are fitted curves.

**Anisotropy in GaTe**. **Figure 3a** shows 3D and side view of the crystal structure of hexagonal and monoclinic GaTe. The synthesized GaTe has a monoclinic phase as proved by HRTEM in **Figure S3**, and the monoclinic phase is found to be more stable than the hexagonal phase[25]. Both phases have a layered structure with two Ga atoms sandwiched between two Te atoms in each layer. But it is clear that the two structures are quite different in the atom arrangements. In the hexagonal phase, each Ga atom is bonded to one Ga atom and three Te atoms and vice versa. All the Ga-Ga bonds are parallel to each other with an identical bonding length of 2.47 Å. However, in the monoclinic phase, every third of the Ga-Ga bonds is twisted and the bonding length is not the same for all Ga-Ga bonds. The Ga-Ga bonding length is 2.48 Å for those labeled by the red rectangle in **Figure 3a** and 2.46 Å for the rest. As a result, a series of atomic chains is formed along the <010> direction. Similar to black phosphorous and $ReS_2$, the formation of these atomic chains is expected to give monoclinic GaTe unique in-plane anisotropy, which is proved by our angle resolved PL and Raman spectroscopy measurements.

**Figure 3b** shows the angle resolved micro-PL (ARMP) and photo reflectance spectra of the GaTe flake. The atomic chains along the <010> direction are clearly seen in the i-FFT HRTEM image, which is parallel to the edge of the bow-tie shaped GaTe flake. The incident laser is polarized along the 0-180º direction and the sample is rotated counter-clockwise from 0º to 360º. It can be seen that the angle resolved PL intensity for both $X_0$ and $X_{sub}$ emission show very similar angular dependence with a twofold symmetry. They reach their maximum intensities when the <010> chain direction is parallel to the laser polarization direction. The maximum PL intensity at 0° is about 5 times larger than at 90°, indicating strong anisotropy of the optical response of the GaTe flake grown on sapphire. This is confirmed in reflectance measurements, which show a minimum along the <010> direction i.e. a maximum in absorbance (assuming negligible transmittance change as a function of rotation angle). This indicates that the maximum PL intensity along the <010> direction is also caused by maximum absorbance of the incident photons, as the emission mechanisms for $X_0$ and $X_{sub}$ are different.

The anisotropic behavior of the monoclinic GaTe flakes is also demonstrated by angle resolved micro-Raman (ARMR) spectroscopy, which has been widely used in other pseudo one dimensional materials such as $ReS_2$[5] and $TiS_3$[9]. The measurement is carried out in the same setup as the ARMP experiment. **Figure 3d** shows the angle resolved Raman intensity of two $A_g$ modes (116 and 212 cm$^{-1}$) and one $B_g$ mode (164 cm$^{-1}$). The anisotropy of the two $A_g$ mode both show a twofold symmetry with a period of 180°. The 116 cm$^{-1}$ mode exhibits maximum intensities along the <010> chain direction, while the 212 cm$^{-1}$ mode is polarized perpendicular to the chain direction. The opposite polarization behavior indicates different atomic displacement directions for the two modes. On the other hand, the polarization of the 164 cm$^{-1}$ mode shows a fourfold symmetry with a 90° period, and the intensity reaches maximum at 45°, 135°, 225° and 315°, making a 45° degree with respect to the <010> chain direction. The different polarization features for the $A_g$ and $B_g$ modes are consistent with recent reports on GaTe anisotropy and can be explained by different symmetries of the $A_g$ and $B_g$ vibration modes[10]. As shown in Figure S4, the GaTe flake is consisted of two domains with different crystal orientation and a grain boundary in between. The SAED pattern shows that each domain is single crystalline with the <010> chains parallel to the edge. Note that the flake in Figure S4 is half the shape of a bowtie shaped GaTe flake as shown in Figure 3c, which is probably due to the damage caused by the sonication when preparing the sample. As a result, the whole flake can be divided into 4 domains indicated by the rectangles in Figure 3c. To further understand the domain structure of the bowtie shaped GaTe flake, we have performed ARMR measurements in different domains on the bowtie shaped flake as shown in Figure S7. The result shows that the 116 cm$^{-1}$ mode is always polarized along the edge direction, which is the <010> chain direction and agrees with the SAED result. The polarization of the 164 cm$^{-1}$ and 212 cm$^{-1}$ modes are also consistent with the crystal orientation in each domain. Thus domain 1 and 3 have the same orientation, whereas domain 2 and 4 have the same orientation. The ARMR result further confirms that the crystal orientation is along the <010> chain direction at the edge of the bow-tie shaped GaTe flake which is consistent with recent work on exfoliated material[10].

Present work establishes the synthesis of anisotropic, direct gap and semiconducting monoclinic GaTe on various substrates including GaAs (111), Si (111) and c-cut sapphire for the first time. Interestingly, nanomaterials grown on sapphire exhibit well defined, narrow, and bright PL emission peaks originating from localized emission due to select type of imperfection sites states that appear at energies well below the fundamental emission line (optical bandgap). Bright emission within the forbidden band is observed for the first time in GaTe and may be the starting point for further defect engineering for optoelectronics in PTMCs. Lastly, angle resolved PL and Raman studies suggest that the synthesized monoclinic GaTe flakes are also highly anisotropic due to its unique crystal structure, which is the first demonstration of the anisotropy in vapor phase synthesized pseudo one-dimensional GaTe.

**Experimental section**

*Sample preparation*: The physical vapor transport (PVT) synthesis of GaTe flakes was carried out in a tube furnace with a 1″ quartz tube. GaTe (60 mg) and $Ga_2Te_3$ (10 mg) powders (American Elements) were mixed together as the source materials and Ar (15 sccm)+$H_2$ (5 sccm) is used as the carrier gas. GaAs (111), Si (111) and c-cut sapphire wafers were used as growth substrates. Prior to growth the GaAs substrates were cleaned by piranha solution followed by dilute HCl solution. Si wafers were cleaned with acetone, methanol, RCA-1 cleanser (mixture of DI water, 27% $NH_3 \cdot H_2O$ and 30% $H_2O_2$ with volume ratio 5:1:1) and 2% HF. C-cut sapphire wafers were cleaned by oxygen plasma for 5 min. The source powders were loaded in a quartz boat and sent to the center of the tube. The substrate was located 10 cm away downstream. The tube was evacuated to 10 mTorr and then heated from room temperature to 650 °C with a ramping rate of 20 °C/min. The temperature was kept at 650 °C for 5 min and then cooled down to room temperature. The Ar flow rate was set at 15 sccm and the growth pressure was 30 Torr for the whole process. The GaTe bulk crystals were synthesized by modified Bridgman growth technique[26] in a single zone furnace at temperatures ranging from 850 °C to 1020 °C for three weeks.

*Materials characterization*: Room temperature PL and Raman measurements for the GaTe flakes were performed in a Renishaw InVia spectroscopy system with a 100x objective lens using a laser source of 488 nm wavelength. The laser was focused onto the sample with a spot diameter of 0.5 µm. Angle resolved measurements were carried out in the same system by mounting samples on a rotation stage and taking data when the sample is rotated every 20º. The incident laser and detector were polarized parallel to each other along the 0-180º direction. Low temperature PL and reflectance measurements were performed in a home build micro-spectroscopy set-up build around a closed-cycle, low vibration attoDry cryostat with a temperature controller (T=4 K to 300K). For PL at a fixed wavelength of 633 nm a HeNe laser was used, for PL experiments at as a function of excitation laser wavelength we used a tunable, continuous wave Ti-Sa Laser SOLSTIS from M SQUARED. The white light source for reflectivity is a halogen lamp with a stabilized power supply. The emitted and/or reflected light was dispersed in a spectrometer and detected by a Si-CCD camera. The morphology and thickness of the GaSe flakes on sapphire were characterized by a Bruker D3100 Scanning Probe Microscope (SPM) in ambient environment. The scanning rate was 1 Hz with a resolution of 512×512. The data was processed by Gwyddion software. Scanning electron microscopy (SEM) was performed on AMRAY 1910 and Hitachi S4700 field emission SEM with a working distance of 12-15 mm and acceleration voltage 15-20 kV. TEM samples were prepared by dispersing the GaTe flakes in isopropyl alcohol (IPA) through sonication and dropping the IPA onto copper grids with holey carbon films.

**Supporting Information**

Supporting Information is attached at the end of this file.

**Acknowledgement**

This work was supported by the Arizona State University seeding program. We gratefully acknowledge the use of facilities at the LeRoy Eyring Center for Solid State Science at Arizona State University. We acknowledge the use of John M. Cowley Center for High Resolution Electron Microscopy at Arizona State University. S.T. acknowledges funding from NSF DMR-1552220. We thank ERC Grant No. 306719, ITN Spin-NANO Marie Sklodowska-Curie grant agreement No 676108 and ANR MoS2ValleyControl, for financial support. X.M. acknowledges Institut Universitaire de France.

Received: ((will be filled in by the editorial staff))
Revised: ((will be filled in by the editorial staff))
Published online: ((will be filled in by the editorial staff))

Supporting Information

**Synthesis of Highly Anisotropic Semiconducting GaTe Nanomaterials and Emerging Properties Enabled by Epitaxy**

Hui Cai, Bin Chen, Gang Wang, Emmanuel Soignard, Afsaneh Khosravi, Marco Manca, Xavier Marie, Shery Chang, Bernhard Urbaszek, and Sefaattin Tongay[*]

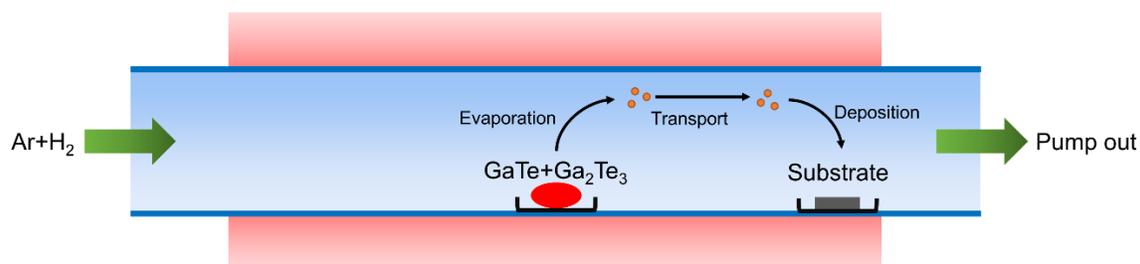

**Figure S1:** Schematic diagram of the PVT system and growth process.

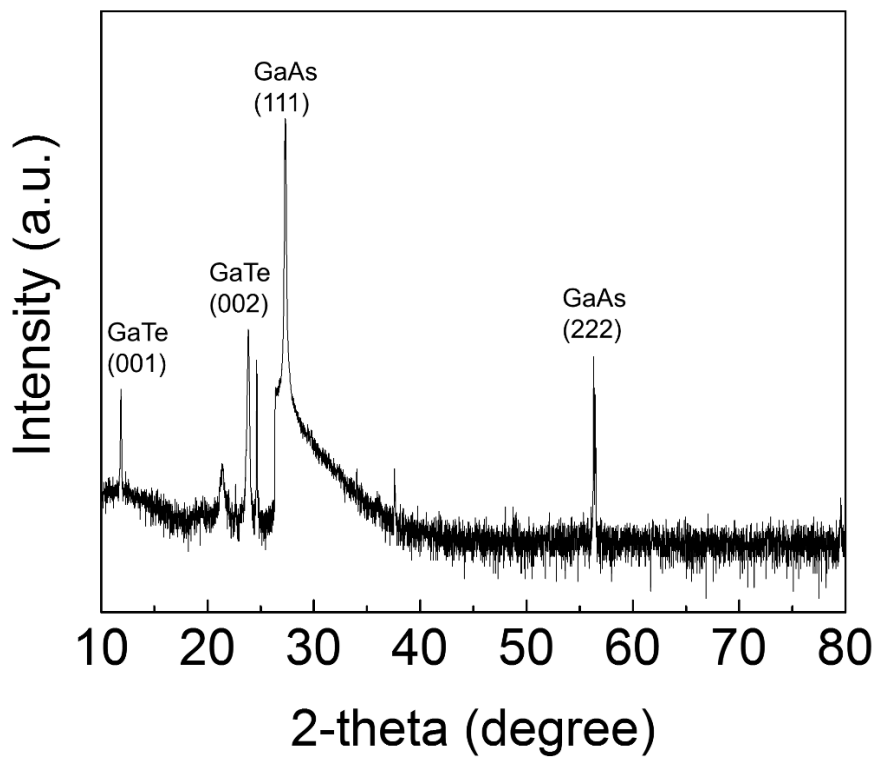

**Figure S2:** XRD spectrum of GaTe grown on GaAs (111).

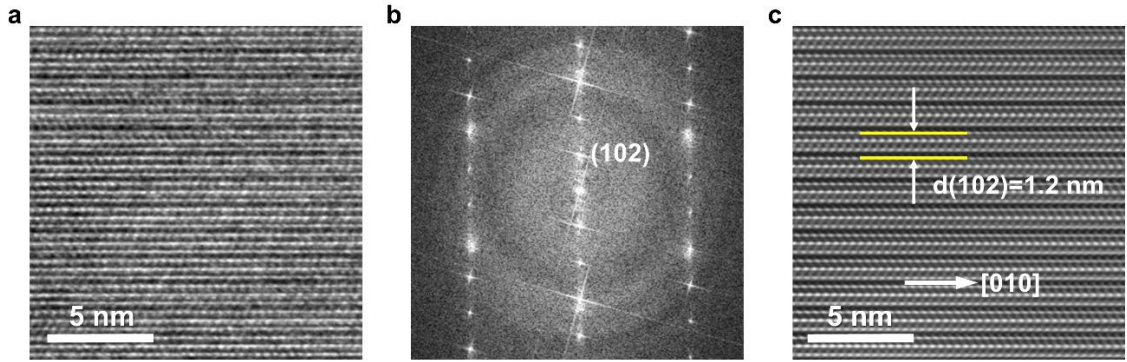

**Figure S3: a).** HRTEM image of the PVT grown GaTe taken along the [$\bar{2}$01] zone axis which is perpendicular to the layers. **b).** FFT image of **(a)** showing the periodicity on the [102] direction. **c).** Inverse FFT image of **(b)** with the [010] chains clearly revealed.

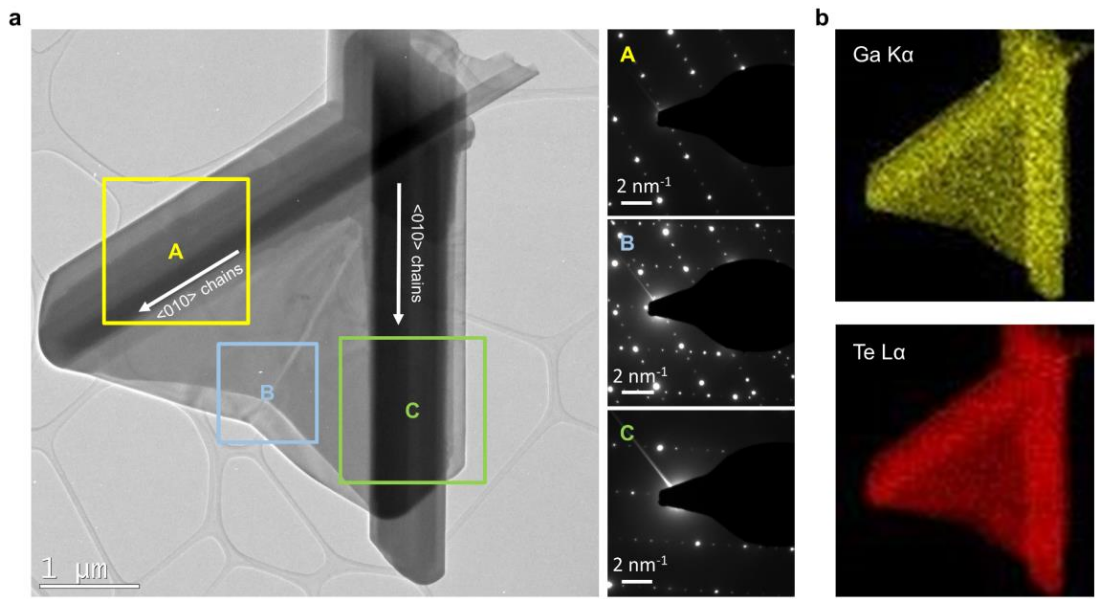

**Figure S4: a).** Low magnification TEM image and selected area electron diffraction (SAED) pattern of a GaTe flake synthesized on sapphire. The SAED patterns are taken at three different regions marked by the squares in the TEM image. **b).** EDS mapping of Ga and Te taken on the GaTe flake.

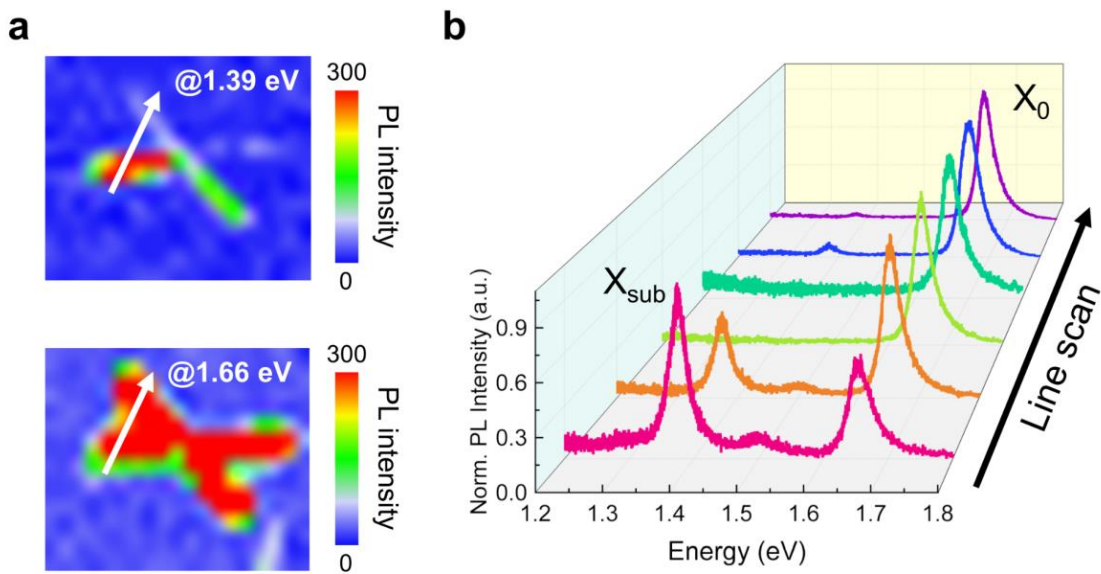

**Figure S5: a).** PL intensity mapping at 1.39 eV and 1.66 eV for flake #2 in Figure 2d, with the arrow indicating the line scan direction. **b).** Line scan of the PL spectra taken at flake #2.

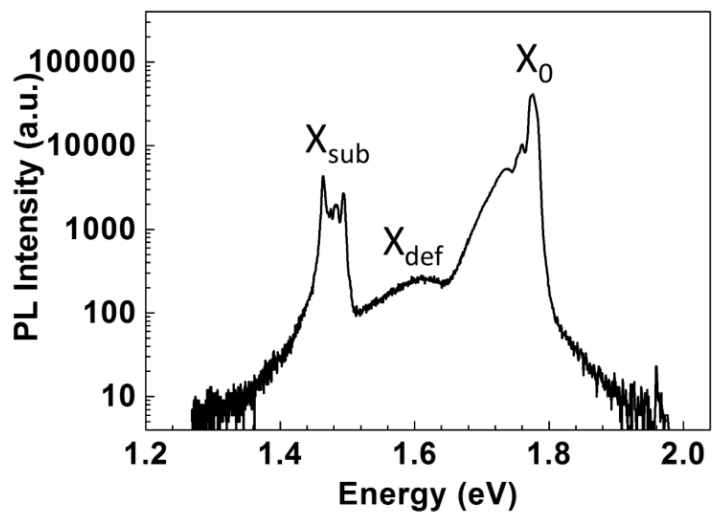

**Figure S6:** PL spectrum measured at 4 K of a GaTe flake synthesized on sapphire (re-plotted from Fig. 2h in log-scale).

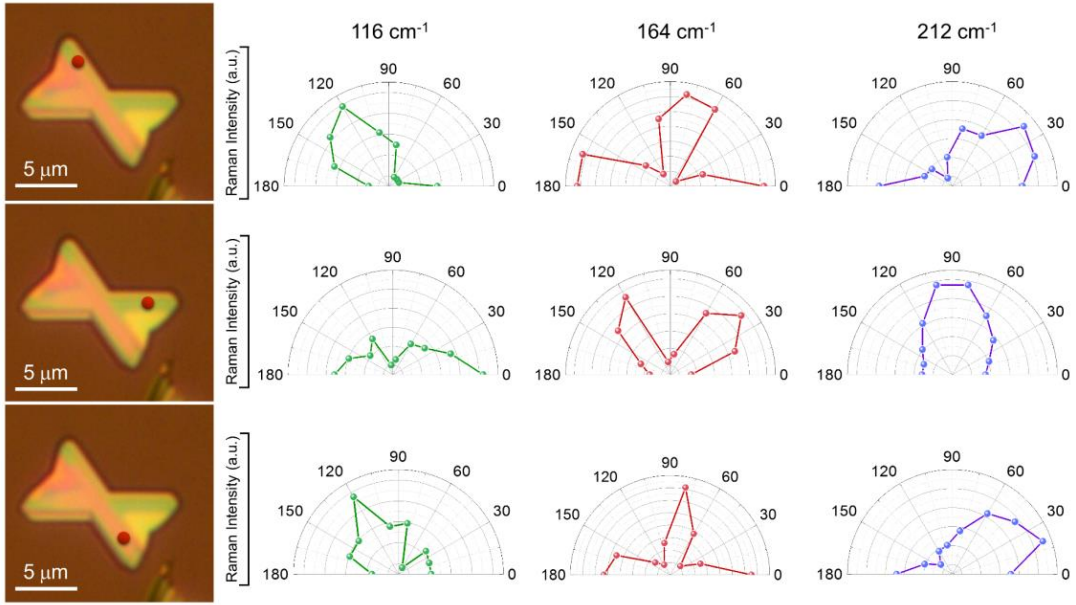

**Figure S7:** Angle resolved Raman spectra of the GaTe flake synthesized on sapphire, taken at 3 different domains as labeled by the red dots.